\documentclass[onecolumn,nobibnotes,nofootinbib,superscriptaddress]{revtex4}
\usepackage{amsmath,amssymb,bm}
\usepackage[a4paper,bindingoffset=0.2in,left=0.8in,right=0.8in,top=1in,bottom=1in,footskip=.25in]{geometry}
\usepackage{graphicx,subfigure,epsfig}
\usepackage{bigints}
\usepackage{color}
\usepackage[breaklinks,colorlinks,urlcolor=blue,citecolor=blue,linkcolor=blue]{hyperref}
\usepackage{mciteplus}
\definecolor{lcolor}{rgb}{0.5,0,0}
\definecolor{citcolor}{rgb}{0,0.3,0.0}
\usepackage[capitalise]{cleveref}
\usepackage{todonotes}
\usepackage{mathrsfs}

\begin{document}

\makeatletter
\newcommand{\correspondingemail}[1]{%
  \begingroup
  \sanitize@url
  \@correspondingemail{#1}%
}
\def\@correspondingemail#1{%
  \endgroup
  \@AF@join{Electronic address:\space\href{mailto:#1}{#1}\space(corresponding author)}%
}
\makeatother

%
\title{Pion and Kaon PDFs via Infrared-Safe Evolution \\
augmented by $J/\psi$ Data Constraints}

\author{Yanbing Cai}
\email{yanbingcai@mail.gufe.edu.cn}
\affiliation{Guizhou Key Laboratory in Physics and Related Areas, Guizhou University of Finance and Economics, Guiyang 550025, China}
\affiliation{Southern Center for Nuclear Science Theory (SCNT), Institute of Modern Physics, Chinese Academy of Sciences, Huizhou 516000, China}
\author{Chengdong Han}
\correspondingemail{chdhan@impcas.ac.cn}
\affiliation{Institute of Modern Physics, Chinese Academy of Sciences, Lanzhou 730000, China}
\affiliation{School of Nuclear Science and Technology, University of Chinese Academy of Sciences, Beijing 100049, China}
\affiliation{State Key Laboratory of Heavy Ion Science and Technology, Institute of Modern Physics, Chinese Academy of Sciences, Lanzhou 730000, China}
\author{Xurong Chen}
\correspondingemail{xchen@impcas.ac.cn}
\affiliation{Southern Center for Nuclear Science Theory (SCNT), Institute of Modern Physics, Chinese Academy of Sciences, Huizhou 516000, China}
\affiliation{Institute of Modern Physics, Chinese Academy of Sciences, Lanzhou 730000, China}
\affiliation{School of Nuclear Science and Technology, University of Chinese Academy of Sciences, Beijing 100049, China}
\affiliation{State Key Laboratory of Heavy Ion Science and Technology, Institute of Modern Physics, Chinese Academy of Sciences, Lanzhou 730000, China}



\begin{abstract}
Probing the partonic structure of the pion and the kaon provides essential insights into the non-perturbative dynamics of QCD, yet their parton distribution functions (PDFs) remain poorly constrained due to the scarcity of high-precision experimental data, especially for the gluon distributions. We present an improved determination of pion and kaon PDFs within the dynamical parton model combined with the maximum entropy method (MEM) framework. Our analysis features two key advancements: firstly, we employ an infrared-safe QCD evolution scheme, allowing the evolution to be reliably extended down to very low $Q^2$, approaching the hadronic scale; secondly, we incorporate pion- and kaon-induced $J/\psi$ hadroproduction data as crucial constraints in the global fit. We find that our approach yields a good description of the available Drell-Yan data, deep-inelastic scattering structure functions ($F_2$), and $J/\psi$ production across various energies and targets. The results provide significantly improved constraints on the gluon distributions at moderate and large $x$ in both the pion and the kaon, offering a more complete picture of their internal structure.
\end{abstract}
\maketitle

\section{Introduction}

Understanding the internal structure of the pion and the kaon, bound states composed of one valence quark and one valence antiquark, remains one of the most fundamental yet challenging problems in hadron physics. As pseudo-Goldstone bosons associated with dynamical chiral symmetry breaking~\cite{Nambu:1960tm,Goldstone:1962es}, the pion and the kaon occupy a unique position in non-perturbative QCD. The pion, being the lightest hadron, plays a central role in mediating the long-range component of the nuclear force. Meanwhile, the kaon, which contains a heavier strange quark, provides a crucial window into the interplay between the emergent hadronic mass generated by QCD dynamics and the explicit symmetry breaking induced by the Higgs mechanism~\cite{Roberts:2021nhw}. Consequently, a precise determination of the parton distribution functions (PDFs) of the pion and the kaon is not only central to understanding the origin of hadron mass but also essential for non-perturbative QCD. However, the PDFs of the pion and the kaon remain poorly constrained in global fits, primarily because of the scarcity of high-precision experimental data. For the pion, existing global fits rely mainly on experimental inputs from pion-induced Drell-Yan processes measured by the NA3 and NA10 collaborations at CERN~\cite{NA3:1983ejh,NA10:1985ibr} and the E615 collaboration at Fermilab~\cite{E615:1989bda}, as well as from leading-neutron deep-inelastic scattering (LN-DIS) at HERA by the H1 and ZEUS collaborations~\cite{H1:2010hym,ZEUS:2002gig}. These measurements provide valuable constraints on the valence and sea quark distributions, while leaving the gluon distributions largely unconstrained. Historically, the only available kaon-induced Drell-Yan data were collected by the NA3 collaboration~\cite{CERN-NA3:1980fhh}, however, these data suffer from limited statistical precision and restricted kinematic coverage ($x>0.2$).

Several QCD analyses of the pion PDFs have been performed using the available experimental data, including GRS~\cite{GRS:1998}, SMRS~\cite{SMRS:1992}, JAM~\cite{Barry:2018}, and xFitter~\cite{Novikov:2020}. These analyses employ diverse theoretical frameworks and methodologies, providing important constraints on the pion PDFs from complementary perspectives. On the theoretical side, notable progress has also been made using Dyson-Schwinger equations (DSE)~\cite{Cui:2020tdf,Bednar:2018mtf}, lattice QCD~\cite{Miller:2025wgr,Gao:2020}, light-front holographic QCD~\cite{deTeramond:2018ecg}, basis light-front quantization~\cite{Lan:2019vui}, and the chiral constituent quark model~\cite{Watanabe:2018}. These theoretical studies provide valuable insight into the non-perturbative dynamics governing meson structure, offering independent benchmarks and constraints that make their predictions complementary to phenomenological extractions from experimental data. A particularly attractive approach for modeling the non-perturbative input of these PDFs is the maximum entropy method (MEM)~\cite{Han:2018wsw,Han:2020vjp,Zhang:2023oja}. In this framework, information-entropy principles are used to constrain the initial parton distributions at the hadronic scale, thereby providing a physically motivated input for subsequent QCD evolution.

This MEM-based strategy has led to particularly simple and predictive initial conditions. In particular, the principle of maximum information entropy~\cite{Han:2020vjp} yields a uniform distribution for the pion's initial valence quarks at the hadronic scale.
Derived solely from entropy maximization under the valence number and momentum sum-rule constraints, this uniform distribution implies that the quark and antiquark are distributed with equal probability over $x\in[0,1]$, consistent with the color-transparency picture of a compact color-dipole state~\cite{Han:2018wsw}. Evolving this non-perturbative input to high $Q^2$ yields valence distributions in excellent agreement with the E615 Drell-Yan data~\cite{Han:2018wsw}. Furthermore, this framework has been successfully extended to a joint analysis of pion and kaon PDFs~\cite{Han:2020vjp}, yielding kaon PDFs consistent with the available NA3 Drell-Yan measurements. However, these pioneering works did not provide comprehensive constraints on the PDFs, particularly for the gluon distributions at intermediate and large $x$. Recently, Chang and collaborators demonstrated that existing pion- and kaon-induced $J/\psi$ production data offer crucial constraints on meson PDFs~\cite{Chang:2020rdy,Chang:2024rbs}. Specifically, these studies found that the cross-section ratios for $J/\psi$ production provide independent evidence that the up valence quark distribution in the kaon is softer than that in the pion, consistent with conclusions drawn from Drell-Yan data. Meanwhile, the $J/\psi$ production ratio is highly sensitive to the gluon distribution in the kaon, serving as a powerful discriminator among existing kaon PDF sets~\cite{Chang:2024rbs}. These findings strongly motivate the inclusion of $J/\psi$ production data in the determination of meson PDFs, as such processes constrain the gluon and sea-quark distributions that are only weakly accessible via Drell-Yan data alone.

A further critical challenge in determining meson PDFs is extending QCD evolution safely into the non-perturbative infrared region ($Q^2 \lesssim 1~\mathrm{GeV}^2$), where the standard massless Dokshitzer-Gribov-Lipatov-Altarelli-Parisi (DGLAP) equations are no longer reliable. Recently, Wang and collaborators~\cite{Wang:2024wny} proposed an infrared-safe evolution scheme that consistently incorporates three key non-perturbative effects: effective parton masses induced by dynamical chiral symmetry breaking, infrared saturation of the strong coupling $\alpha_s(Q^2)$, and parton recombination corrections. This framework enables the evolution to be extended safely down to $Q^2 \approx 0~\mathrm{GeV}^2$, providing a consistent description of PDFs across the entire kinematic range.

In this work, we present an improved determination of the pion and kaon PDFs within the dynamical parton model combined with the MEM framework. The key advances over previous analyses~\cite{Han:2018wsw,Han:2020vjp} are twofold. Firstly, we incorporate pion- and kaon-induced $J/\psi$ production data as additional constraints in the fit, which significantly improves the determination of the gluon distributions. Secondly, we adopt an infrared-safe evolution scheme that allows QCD evolution to be extended safely down to very low $Q^2$. This region includes the hadronic scale in the MEM framework, where the initial PDFs consist solely of valence quarks. The remainder of this paper is organized as follows. In Sec.~\ref{sec_MEM}, we describe the non-perturbative input from the MEM. Sec.~\ref{sec_low_Q2_evolution} presents the infrared-safe evolution scheme with effective mass corrections adopted in this analysis. The color evaporation model (CEM) and $J/\psi$ production are described in Sec.~\ref{sec_jpsi_CEM}. The results are presented and compared with existing experimental data in Sec.~\ref{sec_results}. A summary is provided in Sec.~\ref{sec_summary}.

\section{Initial valence quark distributions from the MEM}
\label{sec_MEM}
The quark model provides the simplest description of meson structure, treating a meson at very low $Q^2$ as being composed solely of a valence quark and a valence antiquark. Indeed, a meson can be regarded as composed exclusively of valence quarks at the hadronic scale, treating the more complex sea quark and gluon components as the result of dynamical evolution. Naturally, parton distributions depend on the resolution scale $Q^2$; by combining the initial input with evolution equations, the PDFs at arbitrary scales can be obtained. The parametrization of the non-perturbative input at the initial low $Q_0^2$ can be given in various functional forms, such as the HERAPDF style~\cite{Sutton:1991ay} or the CTEQ style~\cite{Pumplin:2002vw}. To avoid introducing an excessive number of parameters, we adopt a simple three-parameter form. As described in Ref.~\cite{Han:2018wsw}, this three-parameter form yields results comparable to those obtained from more flexible parametrizations. For the pion, under the assumption of unbroken isospin symmetry and a negligible mass difference between the up and down valence quarks, the valence quark distribution function at the initial scale $Q_0^2$ is expressed as
\begin{equation}
\label{Eq:3par_dis}
u_{\text{v}}(x, Q_{0}^{2})=\bar{d}_{\text{v}}(x, Q_{0}^{2})=A_{\pi}x^{B_{\pi}}\left(1-x\right)^{C_{\pi}}.
\end{equation}
Here, $A_{\pi}$, $B_{\pi}$ and $C_{\pi}$ are undetermined parameters. The exponent $B_{\pi}$ governs the behavior at small $x$ (Regge limit), while $C_{\pi}$ controls the fall-off at large $x$ (threshold region). Furthermore, the normalization constant $A_{\pi}$ ensures the satisfaction of the valence number sum rule
\begin{equation}
\label{Eq:num_sum_pi}
\int\limits_{0}^{1}u_{\text{v}}(x,Q_{0}^{2})dx=\int\limits_{0}^{1}\bar{d}_{\text{v}}(x,Q_{0}^{2})\,dx=1.
\end{equation}
In addition, the parton distribution functions must satisfy momentum conservation
\begin{equation}
\label{Eq:mom_sum_pi}
\int\limits_{0}^{1}x[u_{\text{v}}(x, Q^{2}_0)+\bar{d}_{\text{v}}(x, Q^{2}_0)]\,dx=1.
\end{equation}

It is worth noting that if initial gluon and sea quark components were included, the momentum fraction carried by valence quarks would be less than unity. However, by employing modified evolution equations, the evolution starting scale $Q_0^2$ can be set at a very low point. This scale is low enough that dressed valence quarks can indeed be treated as the sole effective degrees of freedom. As will be shown in Table~\ref{table:1}, the fitted $Q_0^2$ values are indeed sufficiently low ($0.03~\text{GeV}^2$) to justify this valence-only assumption. Therefore, Eq.(\ref{Eq:mom_sum_pi}) is fully justified at the initial low scale.

With the constraints imposed by Eqs.(\ref{Eq:num_sum_pi}) and (\ref{Eq:mom_sum_pi}), two parameters can be fixed, leaving only one free parameter in Eq.(\ref{Eq:3par_dis}). Typically, this parameter is determined by fitting experimental data. However, given the scarcity of data, particularly for the kaon, such fits may not fully capture the constraints across different kinematic regions. This data limitation is a primary reason why our understanding of the internal parton structure of pions and kaons lags behind that of the proton. In recent years, entropy method has found extensive applications in high-energy physics. Various forms of entropy enable an extended characterization of hadron structure beyond classical approaches~\cite{Han:2020vjp,Wang:2014lua,Chen:2024dhz,Hu:2025bla}. Specifically, the MEM provides a conceptually transparent and computationally efficient framework for determining valence quark PDFs, and has been proposed as a viable alternative to conventional global QCD analyses. For the pion and kaon, which are subject to limited theoretical constraints from perturbative QCD, the MEM yields the maximally unbiased PDF parametrization consistent with the available experimental and theoretical information. This approach is rigorously grounded in the principle of maximum entropy, which posits that, given incomplete knowledge of a system, the most appropriate probability distribution is the one that satisfies all known constraints while maximizing the entropy. This ensures maximal objectivity in the face of uncertainty. Accordingly, the entropy of the valence quark distributions at the initial scale can be expressed in terms of the Shannon entropy as follows:
\begin{equation}
\label{Eq:Shannon_entropy}
S=-\int_0^1[u_{\text{v}}(x,Q_0^2)\mathrm{ln}(u_{\text{v}}(x,Q_0^2))+\bar{d}_{\text{v}}(x,Q_0^2)\ln(\bar{d}_{\text{v}}(x,Q_0^2))]dx.
\end{equation}

Using the maximum entropy principle and the constraints from Eqs.(\ref{Eq:num_sum_pi}) and (\ref{Eq:mom_sum_pi}), we find that the non-perturbative inputs satisfy a uniform distribution
 \begin{equation}
 \label{Eq:uniform_dis}
 u_{\text{v}}(x,Q_0^2) = \bar{d}_{\text{v}}(x,Q_0^2) = 1 .
 \end{equation}

Following the same procedure, the initial valence quark distribution functions and constraints for the kaon can also be obtained. The only difference lies in the valence quark flavor content of the pion and the kaon; namely, one only needs to replace the $\bar d$ valence quark in the pion by the $\bar s$ valence quark in the kaon. Once the non-perturbative input is established and the scale $Q^{2}_0$ is fixed, the valence quark distributions can be uniquely determined.

\section{The modified DGLAP equations in an infrared-safe evolution scheme}
\label{sec_low_Q2_evolution}
Within the factorization framework, scattering cross sections require PDFs at arbitrary scales, which are typically obtained using evolution equations. Although the DGLAP equations\cite{Altarelli:1977zs,Gribov:1972ri,Lipatov:1974qm,Dokshitzer:1977sg} are the standard tool for this purpose, they describe linear parton evolution and are most reliable in dilute partonic systems. Since the DGLAP equations account only for parton splitting processes, it fails to capture the dynamics at small $x$, where the gluon density becomes sufficiently large that gluons spatially overlap. Consequently, gluon recombination effects must be taken into account. Mueller and Qiu incorporated these recombination effects, deriving the nonlinear evolution equation in the double leading-logarithmic approximation, known as the Gribov-Levin-Ryskin-Mueller-Qiu (GLR-MQ) equation\cite{Mueller:1985wy}. However, the derivation of the GLR-MQ equation relies on the AGK cutting rules, which leads to an overestimation of the recombination effects and violates momentum conservation. To address this drawback, Zhu, Ruan, and Shen formulated a new evolution equation based on time-ordered perturbation theory (TOPT). This modified DGLAP (MD-DGLAP) equation successfully incorporates antishadowing effects and satisfies momentum conservation\cite{ZhuRuan:1999,zhuw:2007} , providing a more consistent description of parton evolution at small $x$.

To achieve a more physically realistic evolution, the equations must be extended to the extremely low $Q^{2}$ region, as the non-perturbative inputs typically assume that the meson consists solely of valence quarks at a low hadronic scale. Recently, an infrared-safe evolution scheme has been proposed to address this challenge. By incorporating the effects of effective parton masses and a saturating running strong coupling on the basis of parton-parton recombination, this scheme yields evolution equations applicable over the entire $Q^{2}$ region\cite{Wang:2024wny}. Specifically, the flavor non-singlet quark distributions evolve according to
\begin{equation}
\label{Eq:MD-DGLAP-ns}
\frac{d\, x q_i^{\rm NS}}{d \ln Q^2} = \frac{Q^2}{Q^2 + M_q^2}\frac{\alpha_s(Q^2)}{2\pi}\, P_{qq} \otimes x q_i^{\rm NS},
\end{equation}

while the sea quark distributions obey
\begin{equation}
\label{Eq:MD-DGLAP-sea}
\begin{aligned}
\frac{d\, x \bar{q}_i}{d \ln Q^2} &=\frac{Q^2}{Q^2 + M_q^2}\frac{\alpha_s(Q^2)}{2\pi}\left[P_{qq} \otimes x \bar{q}_i + P_{qg} \otimes x g\right] \\
&\quad - \frac{Q^2}{Q^2 + M_q^2}\frac{\alpha_s^2(Q^2)}{4\pi R^2 Q^2}\int_x^{1/2} \frac{dy}{y}\, x P_{gg \to \bar{q}}(x,y)\,[y g(y,Q^2)]^2 \\
&\quad + \frac{Q^2}{Q^2 + M_q^2}\frac{\alpha_s^2(Q^2)}{4\pi R^2 Q^2}\int_{x/2}^{x} \frac{dy}{y}\, x P_{gg \to \bar{q}}(x,y)\,[y g(y,Q^2)]^2,
\end{aligned}
\end{equation}

and the gluon distribution evolves as
\begin{equation}
\label{Eq:MD-DGLAP-g}
\begin{aligned}
\frac{d\, x g}{d \ln Q^2} &=\frac{Q^2}{Q^2 + M_g^2}\frac{\alpha_s(Q^2)}{2\pi}\left[P_{gq} \otimes x \Sigma + P_{gg} \otimes x g\right] \\
&\quad - \frac{Q^2}{Q^2 + M_g^2}\frac{\alpha_s^2(Q^2)}{4\pi R^2 Q^2}\int_x^{1/2} \frac{dy}{y}\, x P_{gg \to g}(x,y)\,[y g(y,Q^2)]^2 \\
&\quad + \frac{Q^2}{Q^2 + M_g^2}\frac{\alpha_s^2(Q^2)}{4\pi R^2 Q^2}\int_{x/2}^{x} \frac{dy}{y}\, x P_{gg \to g}(x,y)\,[y g(y,Q^2)]^2.
\end{aligned}
\end{equation}
The positive term in the third line of Eqs.~(\ref{Eq:MD-DGLAP-sea})-(\ref{Eq:MD-DGLAP-g}) corresponds to the antishadowing correction, which restores momentum conservation that is violated in the original GLR-MQ formulation. In these evolution equations, $P_{qq}$, $P_{qg}$, $P_{gg}$, and $P_{gq}$ denote the standard splitting kernels, while $P_{gg\rightarrow\bar{q}}$ and $P_{gg\rightarrow g}$ are the gluon-gluon recombination kernels~\cite{Wangchen:2017}. $\Sigma$ denotes the quark singlet distribution, defined as the sum over all quark and antiquark flavors. The factor $1/(4\pi R^2)$ in the equations is for two-parton density normalization, with $R$ is the correlation length of two overlapping partons. Compared to the original MD-DGLAP, Eqs.(\ref{Eq:MD-DGLAP-ns})-(\ref{Eq:MD-DGLAP-g}) introduce a new factor $Q^2/(Q^2+M_{q/g}^2)$. This term suppresses the evolution at low $Q^2$, while leaving the evolution unchanged at large $Q^2$, as the factor approaches unity in the regime where $Q^2\gg M^2_{q/g}$. In these evolution equations, $ M_{q/g} $ denotes the mass functions for the quark or gluon, which can be evaluated using continuum Schwinger methods~\cite{Aguilar:2019uob,Roberts:2021nhw}.

To perform the evolution in the infrared region, a crucial ingredient must be properly treated, namely the running coupling constant. As a function of $Q^2$, the one-loop expression of the strong coupling $\alpha_s$ meets divergences at low $Q^2$, known as the Landau pole, which signals the breakdown of perturbation theory in the infrared region. Therefore, an infrared-finite form of the running coupling is required. Based on the concept of the QCD effective charge, a process-independent definition of the strong coupling has been proposed \cite{Roberts:2021nhw,Binosi:2016nme,Cui:2019dwv,Cui:2020tdf}. Within this framework, the $\alpha_s$  saturates to a finite value in the infrared region. This behavior is commonly interpreted as a consequence of the dynamical generation of an effective gluon mass at low momenta, which regulates the infrared divergence. The resulting parametrization of $\alpha_s$ is given by
\begin{equation}
   \alpha_{\rm s} (Q^2)= \frac{\gamma_m\pi}{{\rm ln}[\mathscr{K}^2(Q^2)/\Lambda_{\rm QCD}^2]},
\label{eq:afs}
\end{equation}
with
\begin{equation}
   \mathscr{K}(y) = \frac{a_0^2 + a_1y + y^2}{b_0+y},
\end{equation}
where the parameters are $\gamma_m=4/\beta_0$, $\beta_0=11-(2/3)n_f$, $\Lambda_{\rm QCD}=0.234$ GeV, $a_0=0.104~\text{GeV}^2$, $a_1=0.0975~\text{GeV}^2$, and $b_0=0.121~\text{GeV}^2$.

Given the evolution equations in Eqs. (\ref{Eq:MD-DGLAP-ns})-(\ref{Eq:MD-DGLAP-g}) and the saturation running coupling in Eq. (\ref{eq:afs}), we can perform the parton distribution evolution starting from a low
$Q^2$ scale.

\section{The $J/\psi$ production in the CEM}
\label{sec_jpsi_CEM}
The hadronic production of $J/\psi$ plays a pivotal role in understanding both perturbative and non-perturbative QCD, primarily due to the large mass of the charm quark. Owing to this unique advantage, $J/\psi$ production has attracted sustained experimental and theoretical interest. On the experimental part, extensive data have been recorded for pion- and kaon-induced fixed-target processes, which have greatly advanced our knowledge of their internal structures. In the literature, much of this insight into the internal structure of pion and kaon is derived from the Drell-Yan process, where a quark and an antiquark annihilate to produce a muon pair. Consequently, the Drell-Yan process serves as an excellent tool for constraining the quark structure functions of hadrons. In contrast, the hadronic production of $J/\psi$ is instrumental in constraining the gluon structure functions, as the production mechanism is dominated by gluon-gluon fusion at high energies\cite{Chang:2020rdy}.

On the theoretical part, several models have been proposed to calculate $J/\psi$ production. While these models can describe specific experimental data well within uncertainties, the color singlet model (CSM) and non-relativistic QCD (NRQCD) are the most widely used. These two factorization frameworks share a fundamental similarity in describing $J/\psi$ production: both factorize the production process into short-distance process and long-distance process. In the short-distance process, the $c \bar{c}$ pair is produced, which can be calculated exactly order-by-order using perturbative QCD. Conversely, the long-distance part describes the hadronization of the $c\bar{c}$ pair into a charmonium bound state, involving non-perturbative parameters determined by fitting experimental data. Existing literature presents a wide variety of such long-distance matrix elements. Therefore, in this work, we adopt the CEM to describe $J/\psi$ production, since it requires only one non-perturbative parameter $F$ (the hadronization probability) to be determined from data, while still providing a robust description of fixed-target experimental data.

For pion- and kaon-induced hadronic production of $J/\psi$, the experimental observables are usually presented as functions of Feynman-$x$ ($x_F$). In the CEM, the differential cross section $d\sigma/dx_F$ can be written as
\begin{equation}
\label{Eq:xf_dis}
\begin{aligned}
\frac{d\sigma}{dx_F}=& F \sum\limits_{i,j=q, \bar{q}, g} \int_{2 m_c} ^{2 m_{D}} dM_{c\bar{c}} \frac{2M_{c\bar{c}}}{s\sqrt{x_F^2+4{M_{c\bar{c}}}^2/s}} \nonumber \\
 \times f^{A}_{i}(x_1, \mu_{F}) & f^{B}_{j}(x_2, \mu_{F}) \hat{\sigma}_{ij \rightarrow c\bar{c} X}(x_1, x_2, \mu_{F},\mu_{R}),
\end{aligned}
\end{equation}
with
\begin{equation}
\label{Eq:xf}
  x_F = 2 p_L/\sqrt{s},\qquad x_{1,2} =  \frac{\sqrt{x_F^2+4{M_{c\bar{c}}}^2/s} \pm x_F}{2}.
\end{equation}
Here, $F$ denotes the probability for the hadronization of $c\bar{c}$ pairs into the final $J/\psi$ state. The parameters $m_c$, $m_D$, and $M_{c\bar{c}}$ represent the masses of the charm quark, the $D$ meson, and the $c\bar{c}$ pair, respectively. The terms $f_i^{A}$ and $f_j^{B}$ correspond to the parton distribution functions (PDFs) of the beam pion (or kaon) and the target, respectively. The variables $s$ and $p_L$  are the square of the center-of-mass energy and the longitudinal momentum of the $c\bar{c}$ pair. Furthermore, $\mu_F$ and $\mu_{R}$ denote the factorization and renormalization scales. In our analysis, we adopt $m_c= 1.5~\mathrm{GeV}$, and set the renormalization and factorization scales to $\mu_R=m_c$ and $\mu_F=2m_c$, respectively. In addition, we employ next-to-leading-order (NLO) calculations for $c\bar{c}$ production to obtain more precise and reliable theoretical predictions~\cite{Mangano:1992kq,Nason:1989zy,Nason:1987xz}.

\section{Results and discussions}
\label{sec_results}
Within the MEM framework, the valence quark distribution is assumed to be uniform at the initial scale $Q_0^2$. Since $Q_0^2$ is model dependent, it is treated as a free parameter. Parton distributions at arbitrary scales are then obtained using an infrared-safe evolution scheme. In the evolution equations, $R$ denotes the correlation length between two overlapping partons; as an undetermined parameter, it is also treated as a free parameter. To comprehensively determine the parton distributions of pion and kaon mesons, $J/\psi$ hadronic production data are incorporated to constrain the gluon and sea-quark distributions. As noted previously, calculating the $J/\psi$ cross section involves an undetermined long-distance factor, $F$, which serves as a normalization parameter. Following established practice in the literature, we assume that $F$ is universal across different subprocesses and for both pion- and kaon-induced reactions, allowing only for energy-dependent normalization factors. The value of $F$ is determined by matching the pion-induced $J/\psi$ hadronic production data. The optimized values of $Q_0^2$ and $R$ from our global fit are presented in Table~\ref{table:1}, where the quoted errors correspond to $3\sigma$ uncertainties.

\begin{table}[h!]
  \begin{center}
  \caption{The obtained input scale $Q_0^2$ and $\chi^{2}/\text{ndf}$ results from the fit to experimental data.}
  \begin{tabular}{ccccccccc}
  \hline
  &    & $Q_0^2/\text{GeV}^2$ & $R/\text{GeV}^{-1}$ & $\chi^{2}/\text{ndf}(\text{Drell-Yan+LN-DIS}$) & $\chi^{2}/\text{ndf}(J/\psi)$ \\
  \hline
  & $\pi$    & 0.036(4)   & 3.5(3)   & 1.74 & 5.60  \\

  & $K$    & 0.028(7)   & 2.2(2)  & 0.89  & 1.99  \\
  \hline
    \end{tabular}%
  \label{table:1}
  \end{center}
\end{table}

\begin{figure}[h!]
\begin{center}
\includegraphics[width=0.45\textwidth]{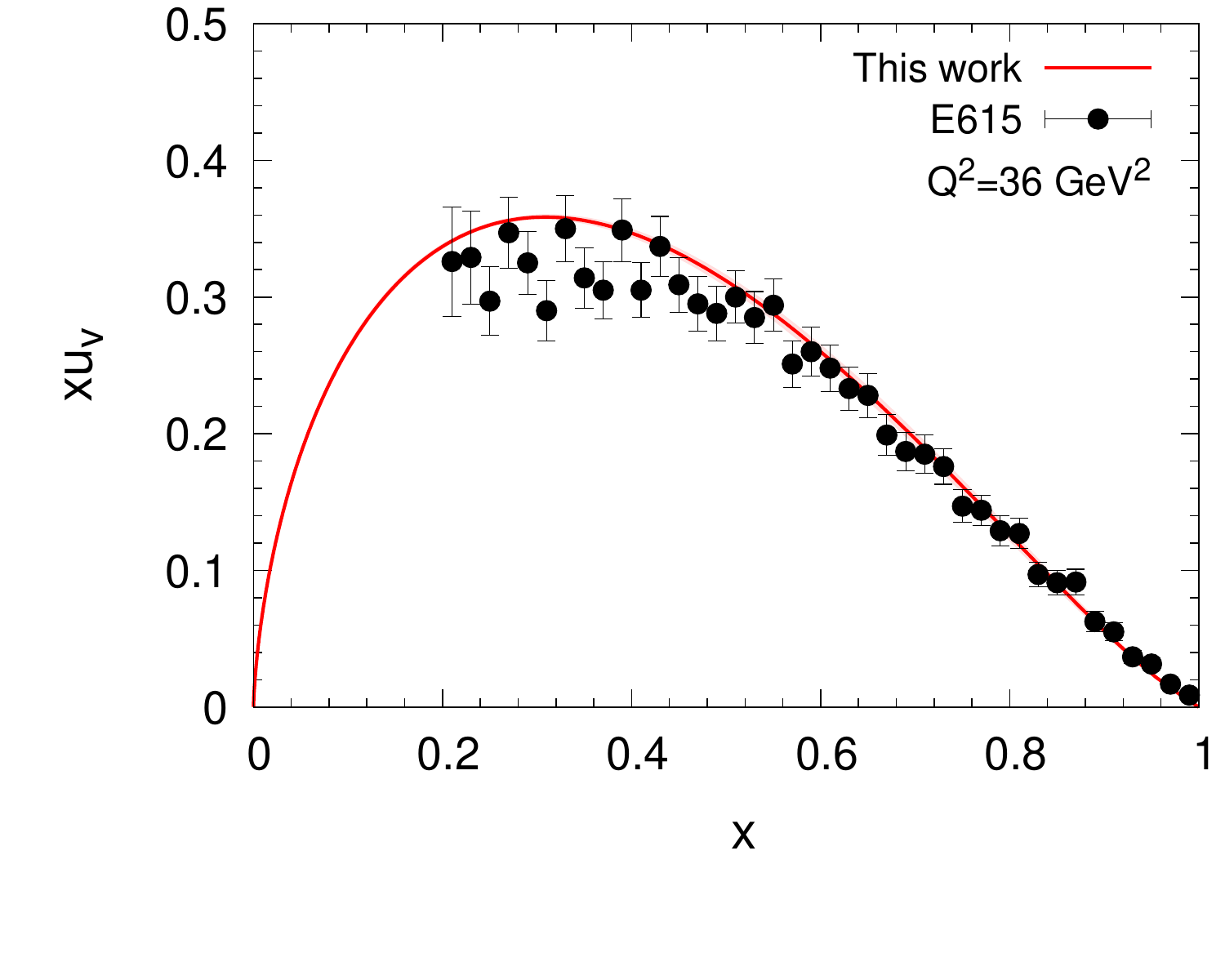}
\vspace{-0.9cm}
\end{center}
\caption{The up valence quark distributions in the pion as a function of $x$. The data are from the E615 experiment \cite{E615:1989bda}.}
\label{fig:uv}
\end{figure}

Figure~\ref{fig:uv} presents the up valence quark distributions in the pion as a function of the momentum fraction $x$. The Drell-Yan experimental data for the pion distribution are taken from the Fermilab E615 experiment\cite{E615:1989bda}. The red solid curve denotes our prediction, and the pink shaded band represents the uncertainty arising from $Q_0^2$ and $R$ (the same hereafter). These Drell-Yan data provide a strong constraint on the up quark distribution from moderate to large $x$, and our results show good agreement with the experimental measurements.

\begin{figure}[h!]
\begin{center}
\includegraphics[width=1.0\textwidth]{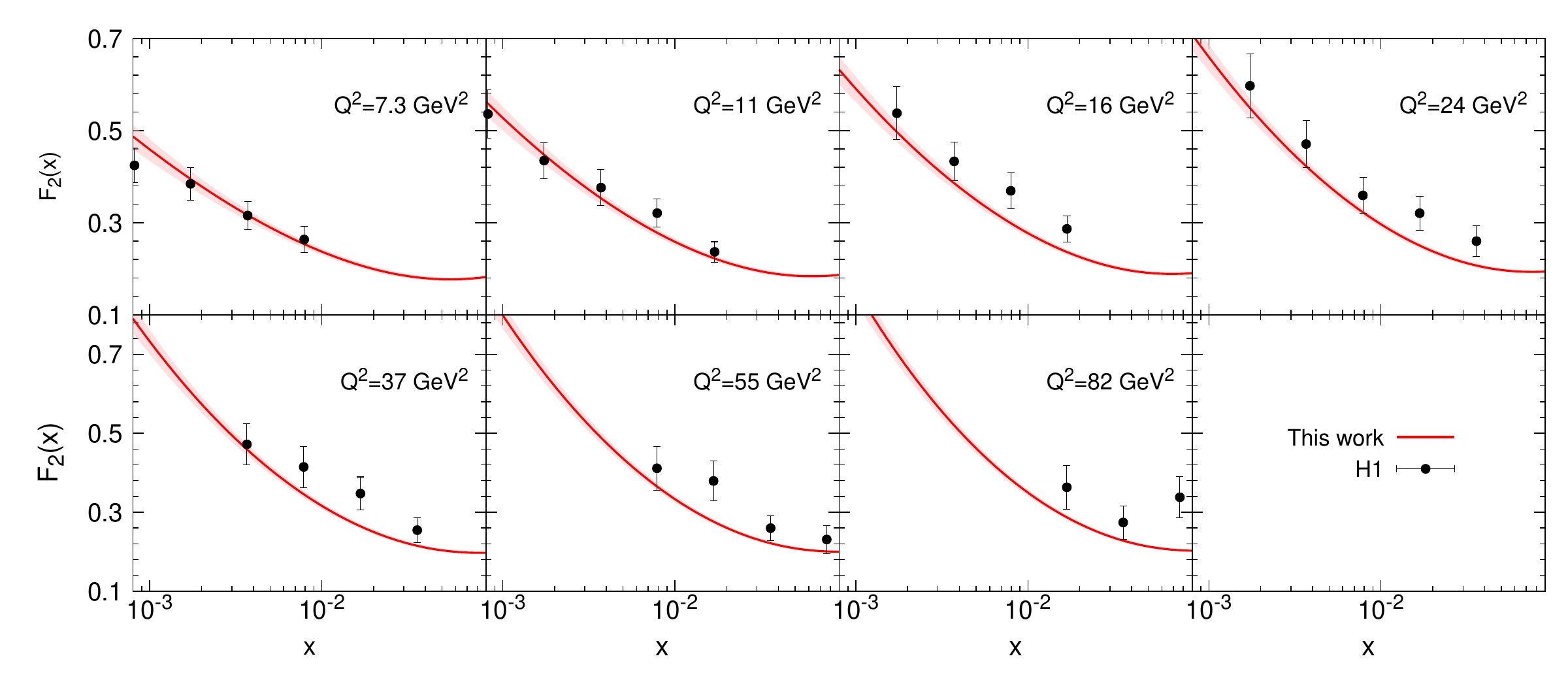}

\end{center}
\vspace{-0.5cm}
\caption{The structure function $F_2(x,Q^2)$ of the pion. The data are from H1 Collaboration at HERA \cite{H1:2010hym}. }
\label{fig:F2}
\end{figure}

As shown in Fig.~\ref{fig:uv}, the Drell-Yan data from the E615 experiment constrain the valence-quark distribution primarily in the moderate-to-large-$x$ region ($x \gtrsim 0.2$). However, these data alone are insufficient for a complete determination of the partonic structure, since the low-$x$ region lacks the necessary constraints on the gluon and sea-quark distributions. To bridge this gap, it is essential to incorporate data that are sensitive to the small-$x$ regime. The structure function $F_2(x,Q^2)$ is directly related to the PDFs, being expressed as a sum of the quark distributions weighted by the squared quark charges. This relation enables the extraction of the pion structure function from LN-DIS experimental data and can therefore be used to constrain the pion PDFs. Figure~\ref{fig:F2} shows the pion structure function $F_2(x,Q^2)$ obtained from the MEM in comparison with data from the H1 Collaboration at HERA~\cite{H1:2010hym}. These LN-DIS data provide direct constraints on the pion structure function down to very small values of $x \sim 10^{-3}$. It should be noted that the ZEUS Collaboration has also measured the pion structure via LN-DIS; however, the extraction of the pion flux involves two different theoretical approaches, namely the effective-flux formula and the additive-quark-model formula~\cite{ZEUS:2002gig}. Owing to these model dependencies in the flux evaluation, the ZEUS data are not included in our global fit, whereas the H1 data provide direct constraints on sea quark distributions in the low-$x$ region. As shown in Fig.~\ref{fig:F2}, the MEM results are in reasonable agreement with the experimental data across the measured kinematic range.

To date, global fit analyses of mainstream pion PDFs have typically not included constraints from $J/\psi$ production. However, as discussed in Ref.~\cite{Chang:2020rdy}, pion-induced $J/\psi$ production can provide additional constraints on pion PDFs, particularly on the gluon density in the large-$x$ region. In this work, we incorporate $J/\psi$ production data into the global fit to constrain the pion PDFs from different kinematic regions. The pion-induced experimental data are taken from CERN~\cite{Corden:1980ht,Tzamarias:1990ij} and Fermilab~\cite{NA3:1983ltt,McEwen:1982fe,E705:1992jno,E672:1995won} experiments. These data are selected because they provide large-$x_F$ coverage for either hydrogen or light nuclear targets (lithium and beryllium) in order to minimize nuclear environment effects, following Ref.~\cite{Chang:2020rdy}. Given the availability of high-precision nuclear PDFs, the CT14nlo~\cite{Dulat:2015mca} parameterization is employed for the hydrogen target in this work, whereas EPPS16~\cite{Eskola:2016oht} is adopted for the lithium, beryllium, tungsten, and platinum targets.

\begin{figure}[t!]
\begin{center}
\includegraphics[width=1.0\textwidth]{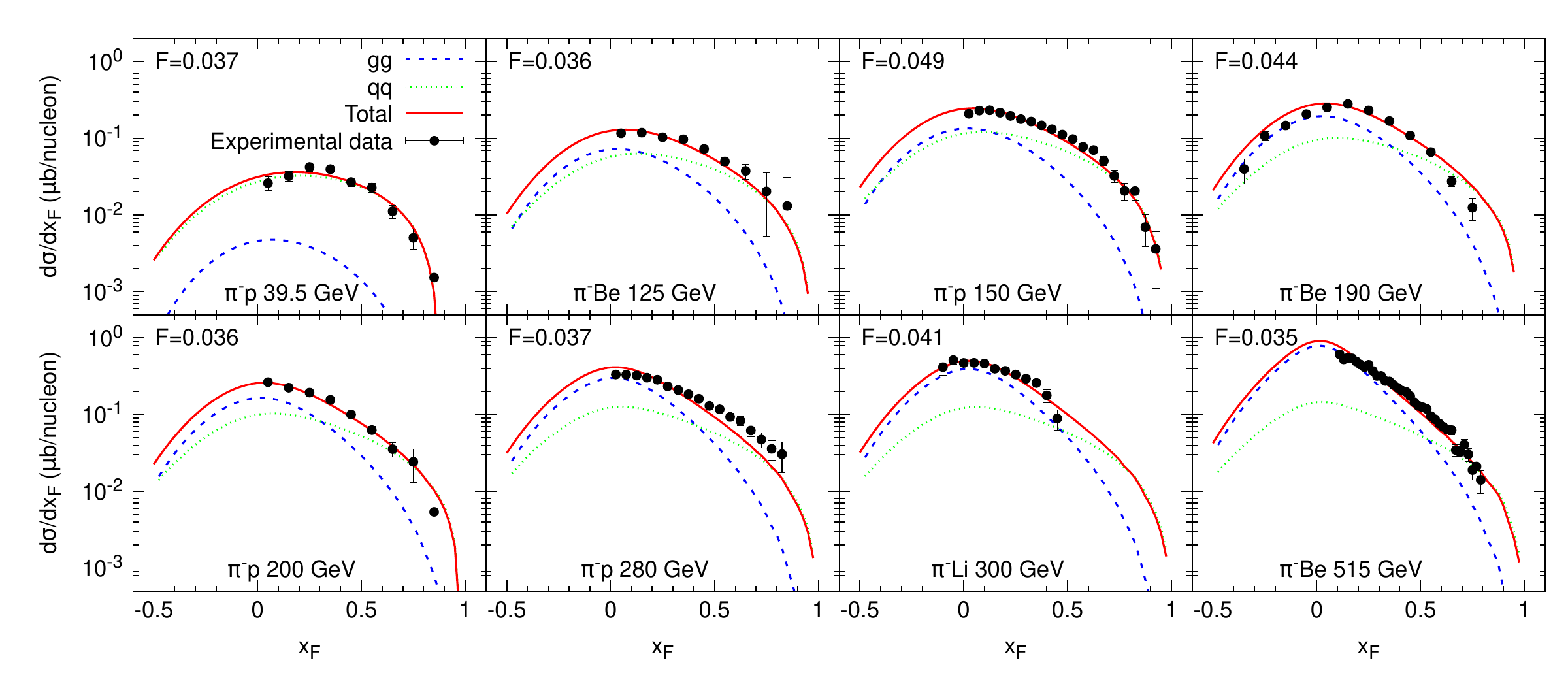}

\end{center}
\vspace{-0.5cm}
\caption{Comparison of the NLO CEM results for $J/\psi$ production off different targets. The data at 39.5 GeV on a hydrogen target are from the CERN WA39 experiment~\cite{Corden:1980ht}. The data at 125~GeV beryllium-target are from the Fermilab E537 experiment~\cite{Tzamarias:1990ij}. The data at 150~GeV hydrogen-target are from the CERN NA3 experiment~\cite{NA3:1983ltt}. The data at 190~GeV beryllium-target are from the CERN WA11 experiment~\cite{McEwen:1982fe}. The data at 200~GeV beryllium-target are from the CERN NA3 experiment~\cite{NA3:1983ltt}. The data at 280~GeV hydrogen-target are from the CERN NA3 experiment~\cite{NA3:1983ltt}. The data at 300~GeV lithium-target are from the Fermilab E705 experiment~\cite{E705:1992jno}. The data at 515~GeV beryllium-target are from the Fermilab E672/E706 experiment~\cite{E672:1995won}.}
\label{fig:jpsi}
\end{figure}

Figure~\ref{fig:jpsi} shows the comparison between the NLO CEM results and the experimental data for $J/\psi$ production off different targets. As can be seen from Fig.~\ref{fig:jpsi}, our results are in good agreement with the experimental data for different nuclear targets and beam energies. In the CEM calculations, the subprocesses responsible for $c\bar{c}$-pair production include the $gg$, $q\bar{q}$, and $qg$ channels. Since the $qg$ process gives a negative contribution that is negligible compared with those from the $gg$ and $q\bar{q}$ channels, only the contributions from the $gg$ and $q\bar{q}$ channels to the total cross section are displayed in the figure. The results for $J/\psi$ production exhibit three characteristic features. First, at the low beam energy of 39.5~GeV, the $q\bar{q}$ channel dominates the cross section over the entire $x_F$ region. Second, as the beam energy increases, the contribution from the $gg$ fusion channel gradually grows around $x_F \sim 0$, whereas it decreases rapidly in the large-$x_F$ region. Third, the $q\bar{q}$ channel provides a relatively flat contribution. At high energies, a crossover between the $gg$ and $q\bar{q}$ channels emerges. The crossover point shifts from $x_F \sim 0.15$ to larger values, reaching approximately $x_F \sim 0.55$. Beyond this point, the $q\bar{q}$ channel becomes dominant again at large $x_F$. From a kinematic perspective, at high energies, the parton momentum fractions around $x_F = 0$ are approximately $0.1$-$0.2$, where gluons dominate and the cross section is highly sensitive to the gluon distribution. In contrast, in the large-$x_F$ region, one parton momentum fraction approaches unity while the other decreases to approximately $0.01$--$0.02$, a kinematic regime in which the quark channel dominates the cross section. These features demonstrate that $J/\psi$ production can provide strong constraints on the pion gluon distribution, particularly in the $x>0.1$ region, which is precisely the kinematic domain where $F_2$ structure-function data are scarce.

The successful constraints on the pion gluon density at large $x$ from $J/\psi$ production demonstrate the unique role of incorporating heavy-quarkonium data into global fits. It is therefore natural and compelling to extend this framework to the study of kaon PDFs, since experimental data for the kaon are even scarcer than those for the pion, particularly with respect to constraints on the gluon distribution. In this work, we apply the same global fit strategy to the kaon, using the available experimental data to constrain its PDFs and to examine the impact of the strange-quark mass on both the gluon and quark distributions.

\begin{figure}[h!]
\begin{center}
\includegraphics[width=0.45\textwidth]{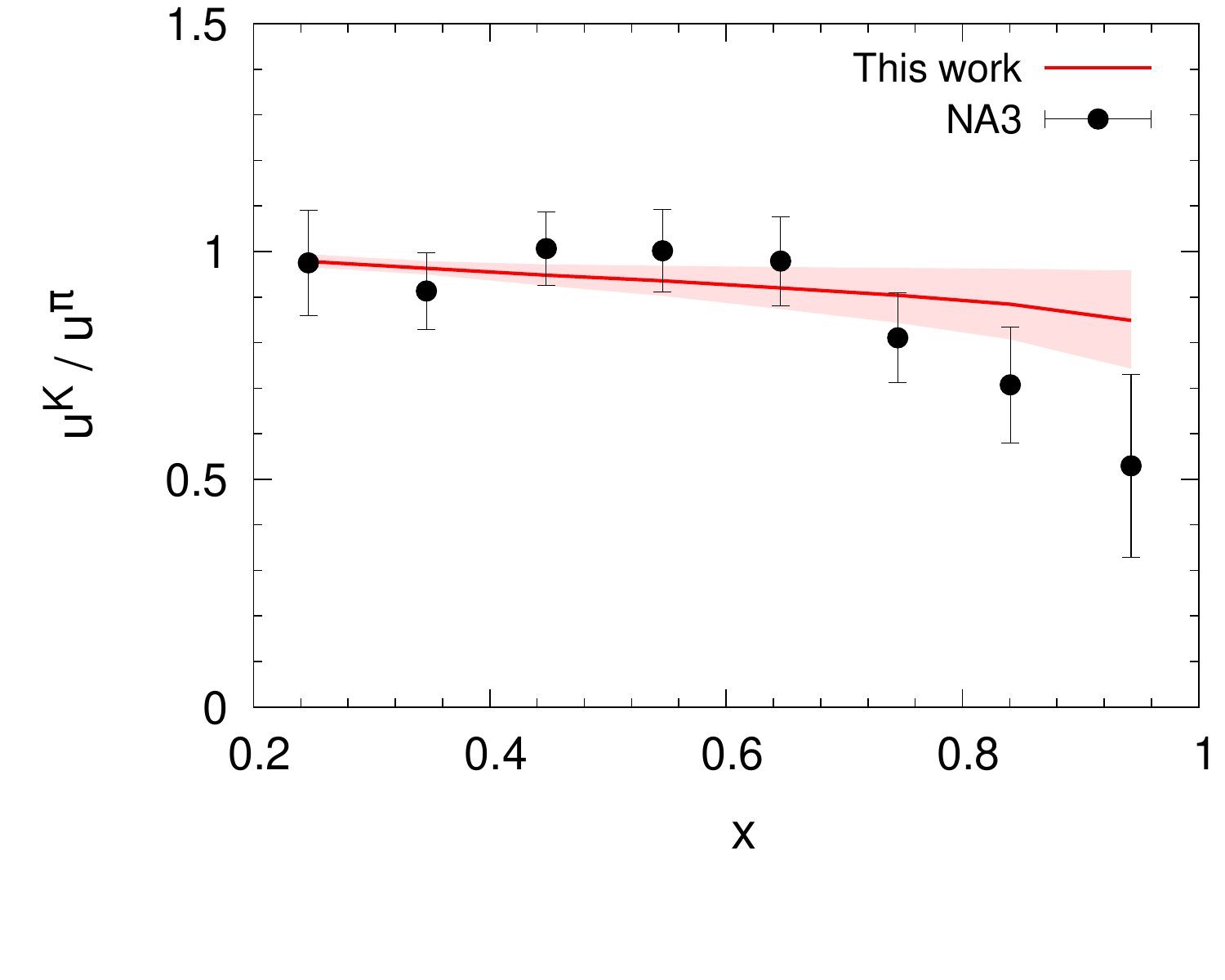}
\vspace{-0.9cm}
\end{center}
\caption{The up quark distribution ratio $u^K/u^{\pi}$ as a function of $x$. The ratio data are from the CERN-NA3 experiment \cite{CERN-NA3:1980fhh}.}
\label{fig:u_ratio}
\end{figure}

For kaons, whose valence constituents are an up quark and a strange quark, the significantly larger mass of the strange quark compared with those of the up and down quarks leads to parton distributions that differ from those of pions. Specifically, valence strange quarks radiate fewer gluons than valence up quarks, and gluon splitting produces fewer $s\bar{s}$ pairs than light $u\bar{u}$ and $d\bar{d}$ pairs. Therefore, following the approach proposed in Ref.~\cite{Cui:2020dlm}, we employ mass-dependent splitting kernels to obtain the high-$Q^2$ kaon PDFs. Figure~\ref{fig:u_ratio} presents the up quark distribution ratio $u^K/u^{\pi}$ as a function of the momentum fraction $x$, with Drell-Yan experimental data taken from the CERN-NA3 experiment~\cite{CERN-NA3:1980fhh}. As can be seen from the figure, this ratio falls below unity in the region $x>0.2$, and the decreasing trend becomes more pronounced as $x$ increases. This behavior indicates that the up quark distribution in the kaon is smaller than that in the pion, which can be attributed to the larger mass of the strange quark.

\begin{figure}[t!]
\begin{center}
\includegraphics[width=1.0\textwidth]{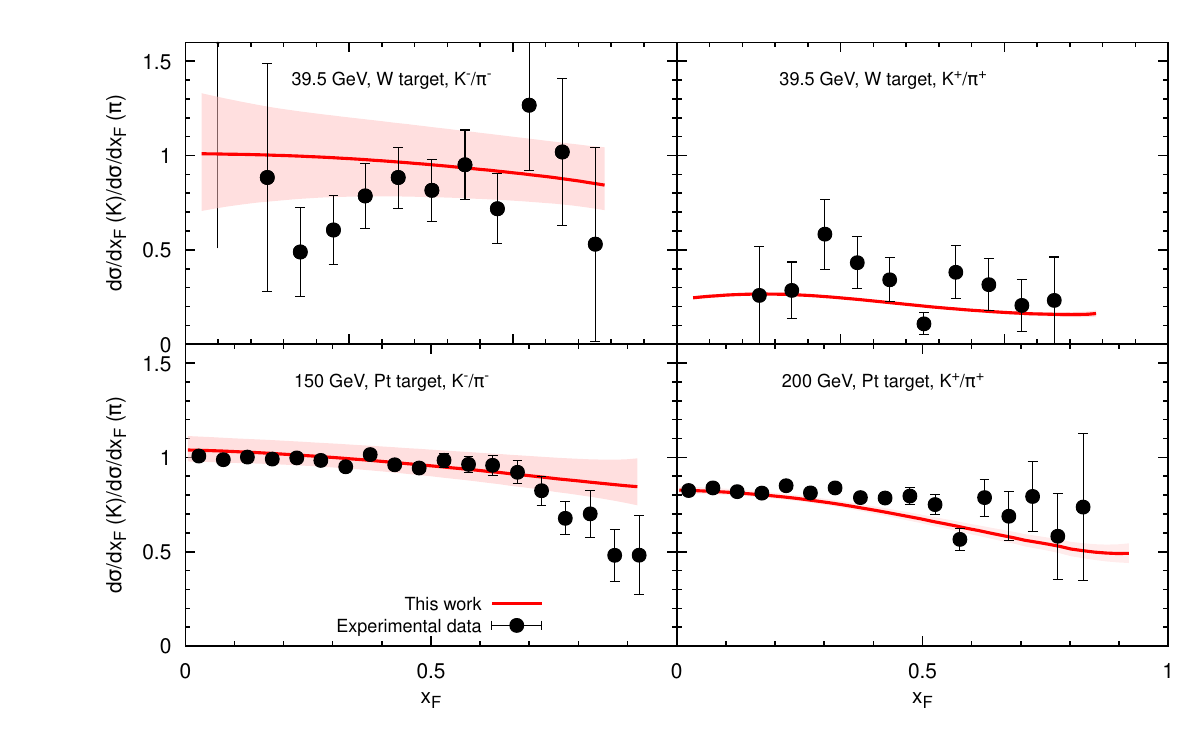}
\hspace{-0.1cm}

\end{center}
\vspace{-0.5cm}
\caption{Ratios of kaon-induced to pion-induced cross section for $J/\psi$ production as functions of $x_F$: $K^-/\pi^-$ at 150~GeV on a platinum target~\cite{NA3:1983ltt}, $K^+/\pi^+$ at 200~GeV on a platinum target~\cite{NA3:1983ltt}, $K^-/\pi^-$ at 39.5~GeV on a tungsten target~\cite{Corden:1980rb}, and $K^+/\pi^+$ at 39.5~GeV on a tungsten target~\cite{Corden:1980rb}.}
\label{fig:jpsi_ratio}
\end{figure}

The $u^K/u^{\pi}$ ratio extracted from Drell-Yan processes currently provides the only available experimental constraint on kaon PDFs in mainstream PDF sets. Moreover, this constraint is restricted to the large-$x$ region ($x>0.2$). In a recent study, the ratios of kaon-induced to pion-induced cross section for $J/\psi$ production were used to analyze kaon PDFs, revealing that the $J/\psi$ production ratio is sensitive to the gluon distribution~\cite{Chang:2024rbs}. Figure~\ref{fig:jpsi_ratio} shows the cross-section ratio data for $J/\psi$ production as a function of $x_F$. Notably, the normalization factor $F$ cancels in the ratio, since it is identical for pion- and kaon-induced production. In our calculations, the nuclear PDFs for tungsten and platinum are taken from EPPS16~\cite{Eskola:2016oht}.

For the $K^-/\pi^-$ ratio, the $J/\psi$ production data at both 39.5~GeV and 150~GeV exhibit a similar trend: the ratio remains close to unity in the small-$x_F$ region and decreases only at large $x_F$. This behavior arises because both $K^-$ and $\pi^-$ contain a valence $\bar{u}$ quark, which can annihilate with the valence $u$ quark in the nuclear target through the $q\bar{q}$ channel to produce a $c\bar{c}$ pair. The suppression of the ratio at large $x_F$ is attributed to the softer valence $\bar{u}$ distribution in the kaon compared with that in the pion, as shown in Fig.~\ref{fig:u_ratio}. In contrast, the $K^+/\pi^+$ ratio data are significantly suppressed. This is because the $K^+$ contains a valence $\bar{s}$ antiquark, which can annihilate only with the sea $s$ quark in the target, whereas the $\pi^+$ contains a valence $\bar{d}$ antiquark that can annihilate with the valence $d$ quark in the target. Consequently, the $K^+/\pi^+$ ratio is strongly suppressed. This effect is particularly pronounced at 39.5~GeV, where the $q\bar{q}$ annihilation channel dominates over the $gg$ fusion process. At the higher energy of 200~GeV, the suppression of the $K^+/\pi^+$ ratio is less pronounced than in the 39.5~GeV case, owing to the increased contribution from $gg$ fusion. Overall, the incorporation of $J/\psi$ production data allows the kaon and pion PDFs to be well constrained and provides a good description of the ratio data for both $K^-$ and $K^+$ beams.

\section{Summary}
\label{sec_summary}

In summary, we have performed a global QCD analysis to determine the PDFs of the pion and kaon. The primary objective was to address the long-standing challenge of constraining the gluon distributions in these mesons, which are essential for understanding the origin of hadron mass.

The MEM is applied to determine the valence quark distributions at the hadronic scale. This method provides a theoretically robust and unbiased parameterization, predicting uniform distributions for the pion and kaon valence quarks. We adopt a modified DGLAP evolution scheme that is infrared-safe. This allowed us to start the evolution from a very low $Q_0^2$ scale where only valence quarks exist, ensuring consistency across the entire kinematic range.  Moreover, we include experimental data on $J/\psi$ production to constrain the gluon distributions. The results show good agreement with Fermilab E615 and CERN NA3 Drell-Yan data (constraining valence quarks), HERA H1 leading-neutron DIS data (constraining small-$x$ sea quarks), CERN and Fermilab $J/\psi$ hadroproduction data (constraining moderate-to-large-$x$ gluon). Our findings confirm that the up valence quark distribution in the kaon is softer than that in the pion. In addition, we provide new and tighter constraints on the gluon distributions in both mesons, particularly in the large-$x$ region.

\begin{acknowledgments}
We are very grateful for the valuable discussion and communication with Wen-Chen Chang. This work has been supported by the Gansu Provincial Youth Talent Program, Grant No. 2026QNGR003, the National Natural Science Foundation of China (Grant Nos. 12305127 and 12547118), the Research Program of State Key Laboratory of Heavy Ion Science and Technology, Institute of Modern Physics, Chinese Academy of Sciences (Grant No. HIST2025CS08), and the National Key R$\&$D Program of China (Grant Nos. 2024YFE0109800 and 2024YFE0109802).
\end{acknowledgments}



\end{document}